\documentclass[11pt,twoside]{article}

\usepackage{asp2006}
\usepackage{epsf}
\usepackage{psfig}
\usepackage{lscape}
\usepackage[dvips]{graphicx}

\begin{document}
\title{Pathways Towards Habitable Moons}   
\author{D M. Kipping$^{1,2}$, S J. Fossey$^{2,3}$, G. Campanella$^{4}$, J. Schneider$^{5}$, G. Tinetti$^{2}$}   
\affil{$^{1}$ Harvard-Smithsonian Center for Astrophysics; $^{2}$ University College London; $^{3}$ University of London Observatory; $^{4}$ Queen Mary, University of London; $^{5}$ LUTH, Observatoire de Paris}    

\begin{abstract} 
The search for life outside of the Solar System should not be restricted to exclusively planetary bodies; large moons of extrasolar planets may also be common habitable environments throughout the Galaxy.  Extrasolar moons, or exomoons, may be detected through transit timing effects induced onto the host planet as a result of mutual gravitational interaction.  In particular, transit timing variations (TTV) and transit duration variations (TDV) are predicted to produce a unique exomoon signature, which is not only easily distinguished from other gravitational perturbations, but also provides both the period and mass of an exomoon.  Using these timing effects, photometry greater or equal to that of the \emph{Kepler Mission} is readily able to detect habitable-zone exomoons down to 0.2 $M_{\oplus}$ and could survey up to 25,000 stars for Earth-mass satellites.  We discuss future possibilities for spectral retrieval of such bodies and show that transmission spectroscopy with JWST should be able to detect molecular species with $\sim 30$ transit events, in the best cases.
\end{abstract}


\section{Introduction}   

In this conference, named `The Pathways Towards Habitable Planets', we are trying to develop a strategy towards detecting life on another planet.  However, by only considering \emph{planets} we may be unnecessarily restrictive by excluding another potentially common environment for life to thrive - moons.  The possibility of life-bearing moons was addressed by Williams et al. (1997) who noted that a moon of around one third of an Earth mass or larger, residing around a gas giant planet in the habitable zone, would satisfy many of the criteria traditionally placed on habitable planets.  For M-dwarf systems, habitable-zone moons have the added bonus that any tidal locking of the planet or the moon would not result in one side of the moon being locked in perpetual darkness.

Whether or not such moons could be dynamically stable remained a relativity open question until Barnes \& O'Brien (2002) showed that a habitable-zone Jupiter-like planet could hold onto a 1$M_{\oplus}$ moon in systems with a host star as small as $M_* \sim 0.4 M_{\odot}$, and remain dynamically stable for 5 Gyr.  In light of these two studies, the potential for habitable moons must be taken as being at least theoretically possible.  The formation of a $\sim 1 M_{\oplus}$ exomoon around a Jupiter-mass planet is not supported by the current models of the formation of the regular satellites, e.g. see Canup \& Ward (2006).  However, the possibility of a Jupiter-like planet capturing a terrestrial planet as a moon (an irregular satellite) is a topic on which we could find no detailed studies. Given that such moons would be dynamically stable, and the lessons learnt from the failings of trying to predict exoplanetary systems based on Solar System models prior to the first actual detection, a search for extrasolar moons is not only historically justified but scientifically imperative.

\section{Transit Timing Effects due to Extrasolar Moons}
\subsection{Transit Timing Variation (TTV)}

Extrasolar moons do not significantly affect the motion of the host star in a way which would be detectable through radial velocity techniques.  It has been proposed that the moons of planets in pulsar systems could be detected through time-of-arrival analysis of the host star's pulses (Lewis et al. 2006), but we can find no publications in the literature proposing an analogous methodology for radial velocity.  A discussion of other proposed methods is briefly given by Kipping et al. (2009c), but here we limit ourselves to transiting exoplanet-moon systems.  Ostensibly, the most obvious detection method in such cases is the occultation of the star light by the moon itself (Sartoretti \& Schneider 1999).  However, the position of this signal will appear at different relative positions to the planet for each transit event, with a mean position located directly over the planetary signal.  Consequently, we would require a fortunate aligment to make the detection.  This point aside, perhaps the most critical objection to detecting exomoons in the same way exoplanets are found in transit surveys, is that planets can be confirmed through radial velocity whereas moons cannot.  CoRoT, HAT-NET and WASP have all reported that the vast majority of their planetary signals are infact not real planets ($\sim 98$\% for CoRoT) due to the plethora of effects which can mimic such signals (see Almenara et al. 2009; Hartman et al. 2004; Pollacco et al. 2006).  Ergo, by the same reasoning, it is imprudent to claim an exomoon can be convincingly detected in this manner.

Sartoretti \& Schneider (1999) were the first to propose that exomoons could be detected around transiting exoplanets by looking for transit timing variations (TTV).  For a planet-moon system, the two bodies orbit a common center of gravity which itself orbits the host star.  Consequently, the planet's tangential position can appear shifted from that expected of a strictly linear ephemeris by a maximal spatial deviation of $a_W = a_S M_S/M_P$, where $M_P$ and $M_S$ are the planetary and satellite masses respectively and $a_S$ is the semi-major axis of the moon's orbit (where we have assumed the simple case of a coplanar, edge-on system with circular orbits).  This spatial deviation is observeable as a timing deviation in the transit lightcurve where $\delta t \sim a_W/v_P$ where $v_P$ is the tangential velocity of the planet.   The effects of planetary and lunar orbital eccentricity are accounted for in the more elaborate derivation of Kipping (2009a) as well reformulating the expression in terms of root-mean-square amplitude to reflect the way such timing signals as proposed to be detected, i.e. by excess scatter.

\begin{equation}
\delta_{TTV} = \frac{1}{\sqrt{2}} \cdot \frac{a_P^{1/2} a_S M_S M_{PRV}^{-1}}{\sqrt{G (M_* + M_{PRV})}} \cdot \frac{\zeta_T(e_s, \varpi_S)}{\Upsilon(e_P,\varpi_P)}
\end{equation}

Where $\zeta_T$ and $\Upsilon$ are terms absorbing the effects the orbital eccentricity, we reference readers to the original paper for details.

\subsection{The Problems with TTV}

Although the TTV amplitude has been derived, how would such a signal be detected?  One critical problem with exomoon timing signals is that due to dynamical constraints, the orbital period of a moon ($P_S$) will always be significantly lower than the orbital period of a planet ($P_P$), but we only see a transit once every planetary period.  Kipping (2009a) showed that $P_S \leq P_P \sqrt{\chi^3/3}$ where $\chi$ is the orbital distance of the satellite in units of Hill radii.  For a prograde exomoon, Domingoes et al. (2006) showed that $\chi_{max} \simeq 0.4895$ meaning $P_S \leq 0.2 P_P$.  Consequently, aliasing is unavoidable and a Fourier analysis will reveal a range of possible harmonic frequencies for the lunar period.

If we are unable to provide a singular solution for $P_S$ (or equivalently $a_S$ by re-arrangement through Kepler's Third Law) then equation (2) has no unique solution.  Thus a range of possible exomoon masses are possible since the TTV amplitude only gives us $M_S \times a_S$.  The problem may be averted if only one harmonic frequency falls within the range of dynamic stability, but short period exomoons will produce numerous possible solutions.  Additionally, we have so far only dealt with two free parameters, period and mass, the addition of another unknown, like inclination, makes a unique solution unfeasible.

Finally, another critical problem was that due to the aliasing of frequencies, the TTV signal could infact appear as a planetary perturber.  Holman \& Murray (2005) showed that planetary bodies could induce TTV on a transiting planet with a range of frequencies.  From the TTV information alone, there is no way to distinguish between the planet or the moon scenario.

\subsection{Transit Duration Variation (TDV)}

To resolve these problems, Kipping (2009a \& 2009b) proposed using transit duration variation (TDV) to break the degeneracy.  In the simple coplanar case, a planet's tangential velocity at the moment of transit appears to oscillate around some local mean value as a result of the moon's presence.  Changes in velocity will directly affect the transit duration, $T$, thus leading to TDV.

It is important to understand the differences between TTV and TDV.  TTV is caused by changes in the planet's position, which is analogous to the astrometry planet-hunting method, where changes in the star's position imply a companion.  TDV is caused by changes in the planet's velocity, which is analogous to the radial velocity planet-hunting method.  Like astrometry, TTV is more sensitive to distant companions and scales as mass multiplied by distance, or $M_S a_S$.  Like radial velocity, TDV is more sensitive to close-in companions and scales as $M_S a_S^{-1/2}$.  Kipping (2009a) showed that:

\begin{equation}
\delta_{TDV} = \sqrt{\frac{a_P}{a_S}} \sqrt{\frac{M_S^2}{M_{PRV} (M_{PRV} + M_*)}} \frac{T}{\sqrt{2}} \frac{\zeta_D(e_s, \varpi_S)}{\Upsilon (e_P,\varpi_P)}
\end{equation}

Where again $\zeta_D$ and $\Upsilon$ are terms absorbing the effects of orbital eccentricity.  The scaling of TDV amplitude to $M_S a_S^{-1/2}$ can be seen within this expression.

\subsection{The Complementary Nature of TTV \& TDV}

TDV and TTV are highly complementary due to their different scalings and different phases.  Just like simple harmonic motion, the position and velocity of the planet have a $\pi/2$ phase difference (for circular orbits) which provides a highly unambiguous signal.  For example, a perturbing planet is not predicted to produce a TDV effect with such a phase difference.  Secondly, TDV allows one to detect close-in exomoons which TTV would miss and vice versa meaning the whole parameter space may now be probed.

Furthermore, the different scalings mean that the ratio of the TTV to the TDV amplitude provides $a_S$ (or equivalently $P_S$) and $M_S$ uniquely.  The derived value of $P_S$ may then be compared to the Fourier spectrum to identify the true orbital frequency and further constrain the system.  Any significant deviation between the Fourier frequency and the ratio-derived frequency would indicate orbital eccentricity or inclination.

\section{Detectability of Habitable-Zone Exomoons}

Despite the theory behind detecting exomoons being now well-established, no detections have been made.  This may be primarily due to the fact all of the known transiting exoplanets are short-period or highly eccentric, which dynamically forbid exomoons on Gyr timescales.  However, with the launch of \emph{Kepler}, we expect that many long-period transiting planets will shortly be announced.  Kipping et al. (2009) evaluated the TTV and TDV signal-to-noise strengths with \emph{Kepler}-class photometry (KCP) for a range of orbital configurations, but fixing the planetary period to always be that of a habitable-zone planet.  This was done to limit the parameter space to a more computationally feasible volume and also because such moons would be the most interesting cases, in light of this conference.  Details of the methods and assumptions employed can be found in Kipping et al. (2009) and we summarize the key results here for brevity.

\begin{itemize}
\item Low-density planets are the ideal hosts for detection as they provide the largest transit depth but experience the greatest perturbation from a companion.  
\item 25,000 stars in \emph{Kepler's} field-of-view are bright enough to be surveyed for a habitable-zone Saturn-like exoplanet hosting an Earth-mass exomoon.  A Galactic plane survey with KCP should expand this survey to around 2 million stars within the detectable range.
\item For the ideal, but realistic, case of a M2V dwarf at 10pc, KCP could detect a habitable-zone exomoon down to 0.2 $M_{\oplus}$.  
\item Saturn-Titan systems are not detectable with KCP but the results do suggest that large exomoons should be found in the coming years should they exist.
\end{itemize}

\section{Characterizing Exomoons}
\subsection{Radius and Internal Structure}

Once such a moon is detected through timing deviations, what more can be learnt of the satellite?  The TTV and TDV effects provide the orbital period of the moon which may then be fitted through the data points to obtain an exomoon ephemeris.  We also have multiple transit lightcurves of the planet-moon system.  For a small moon, the signal will appear like a typical planetary signal and the moon signal will be at the same level as the noise.  However, for each transit event we can subtract the planetary transit signal and then fold the residuals on the exomoon period.  This will produce a composite exomoon transit lightcurve, which should beat down the noise at roughly the reciprocal of the square root of the number of transits observed.

With KCP, let us consider the 0.2 $M_{\oplus}$ habitable moon with an M2V host star. We expect the lunar radius to be 0.65$R_{\oplus}$ based upon a terrestial planet model of Valencia et al. (2006) thus producing a transit depth of $\sim 140 \pm 17$ppm lasting for 50 minutes (using $b=0.5$).  However, with a period of 0.126 years, we would obtain around 28 transits over the \emph{Kepler Mission} lifetime thus reducing the error from a 12\% uncertainty to a 2\% uncertainty (3.2 ppm).  Obtaining the radius to this precision is critical in constraining the internal structure of the exomoon, i.e. distinguishing between icy and rocky (Valencia et al. 2006).

\subsection{Exomoon Transmission Spectroscopy}

In the same way, transmission spectroscopy could be potentially performed.  For the ideal case of an M2V star at 10pc, the apparent magnitude is 9.5 in \emph{Kepler}'s bandpass but is likely to be around 5.5 in K-band for infrared spectroscopy, i.e. approximately as bright as HD 189733.  On this target, Knutson et al. (2007) obtained 8$\mu$m photometry with 0.4 second cadence and rms scatter 4.325 mmag with \emph{Spitzer}.  Over a 50 minute integration this reduces to 49.9 ppm.  JWST will have a collecting area 44 times larger than \emph{Spitzer} and thus we expect this same signal to be reduced to 7.5 ppm per transit, comparable to \emph{Kepler} in the visible. The scale height of a planetary atmosphere is roughly $H = k T/g \mu$. For a terrestrial planet with mass 0.2 $M_{\oplus}$ and radius 0.65 $R_{\oplus}$, we have $H \simeq 20$km.  For a strong molecular transition, the radius of the planet may appear to change by a few scale heights, say 3, which would change the transit depth by $\sim 4$ppm.  This should be detectable to 1-$\sigma$ confidence by binning 4 transits together and to 3-$\sigma$ by binning 32 transits.

%

\subsection{Direct Imaging}

Potentially, the most powerful method for characterising an extrasolar moon would be a spectroscopic investigation through direct imaging.  This would necessitate the ability to resolve the planet and moon contributions. Spatially resolving such signals would require multi-kilometric interferometers (Schneider et al. 2010) and is likely to be several decades away from being realized. However, it will be possible to make spectroscopic observations of the unresolved planet-moon pair with facilities like the Terrestrial Planet Finder Coronagraph (Ford et al. 2004) or the 42m European Extremely Large Telescope (Diericks et al. 2004), both of which are designed with the required sensitivity to detect Earth-sized planets. The combined light could then be resolved by subtracting the planet spectrum from the planet-moon spectrum during `mutual events'; for example planet-moon eclipses or the disappearance of the moon in the shadow of the planet (Cabrera \& Schneider 2007).

\section{Conclusions}

We have shown that exomoons, in particular habitable exomoons, are detectable by coupling transit timing variations (TTV) and transit duration variations (TDV) together down to 0.2 $M_{\oplus}$ for \emph{Kepler}-class photometry.  Around 25,000 stars could be surveyed within \emph{Kepler}'s field-of-view for Earth-mass habitable exomoons.  Indeed, we propose here that it may even be possible for gas giants of $\sim 10 M_J$ to harbour a super-Earth mass exomoon, or a `super exomoon'. Subsequent characterisation of exomoons could be achieved by using the planet-moon ephemeris derived from the transit timing signal.  For the case of a system within 10pc and a 0.2 $M_{\oplus}$ habitable exomoon, we predict that molecular species could be found using transmission spectroscopy with JWST after the binning of $\sim$30 transit events.



\end{document}